\documentclass[fleqn, usenatbib,useAMS]{mnras}

\usepackage{graphicx}
\usepackage{amsmath,amssymb}
\usepackage{color}
\usepackage{hyperref}
\usepackage{multicol}
\usepackage{multirow}
\usepackage{mathtools}
\usepackage{siunitx}
\usepackage[dvipsnames]{xcolor}
\usepackage[left]{lineno}
\usepackage{pdflscape}	
\usepackage{subfigure}

\usepackage[T1]{fontenc}
\usepackage{ae,aecompl}
\usepackage{newtxtext,newtxmath}

\newcommand{\beq}{\begin{equation}}
\newcommand{\eeq}{\end{equation}}
\newcommand{\bdm}{\begin{displaymath}}
\newcommand{\edm}{\end{displaymath}}

\definecolor{Gray}{gray}{0.9}

\newcommand{\redtext}[1]{#1}  
\newcommand{\greentext}[1]{{ #1}}  

\graphicspath{{./plots/}}

\title[Constraining GWs from BNS merger remnants]{Constraining the Gravitational-Wave Afterglow From a Binary Neutron Star Coalescence}

\author[S. Banagiri et al.]{Sharan Banagiri,$^{1}$\thanks{E-mail: banag002@umn.edu} Michael W. Coughlin,$^{2}\thanks{Email: mcoughli@caltech.edu} $James Clark,$^{3}$ Paul D. Lasky,$^{4}$ $^{5}$\newauthor M.~A.~Bizouard,$^{6}$ Colm Talbot,$^{4}$ $^{5}$ Eric Thrane $^{4}$ $^{5}$ and Vuk Mandic$^{1}$ 
\\
$^{1}$School of Physics and Astronomy, University of Minnesota, Minneapolis, Minnesota 55455, USA\\
$^{2}$Division of Physics, Math, and Astronomy, California Institute of Technology, Pasadena, CA 91125, USA\\
$^{3}$Department of Physics, Georgia Institute of Technology, Atlanta, GA 30332, USA\\
$^{4}$School of Physics and Astronomy, Monash University, Clayton, Victoria 3800, Australia\\
$^{5}$OzGrav: The ARC Centre of Excellence for Gravitational-wave Discovery, Monash University, Clayton, Victoria 3800, Australia\\
$^{6}$Artemis, Universit\'e C\^ote d'Azur, Observatoire C\^ote d'Azur, CNRS, CS 34229, F-06304 Nice, France}

\pubyear{2019}

\begin{document}
\label{firstpage}
\pagerange{\pageref{firstpage}--\pageref{lastpage}}
\maketitle

\begin{abstract}
Binary neutron star mergers are rich laboratories for physics, accessible with ground-based interferometric gravitational-wave detectors such as Advanced LIGO and Advanced Virgo. If a neutron star remnant survives the merger, it can emit gravitational waves that might be detectable with the current or next generation detectors. The physics of the long-lived post-merger phase is not well understood and makes modeling difficult. In particular the phase of the gravitational-wave signal is not well modeled. In this paper, we explore methods for using long duration post-merger gravitational-wave signals to constrain the parameters and the properties of the remnant. We develop a phase-agnostic likelihood model which uses only the spectral content for parameter estimation and demonstrate the calculation of a Bayesian upper limit in the absence of a signal. With the millisecond magnetar model, we show that for an event like GW170817, the ellipticity of a long-lived remnant can be constrained to less than about 0.5 in the parameter space used.

\end{abstract}


\section{Introduction}

The detection of gravitational-wave signals from binary black hole mergers \citep{AbEA2016a,AbEA2017c, LIGOScientific:2018mvr}, and the binary neutron star (BNS) merger GW170817~\citep{TheLIGOScientific:2017qsa} by Advanced LIGO and Advanced Virgo~\citep{TheLIGOScientific:2014jea, TheVirgo:2014hva} in their first and second observing runs (O1 and O2) show that compact binary coalescences are primary sources of gravitational waves (GW) for terrestrial gravitational-wave detectors. BNS mergers in particular provide an extremely rich environment for studying physics at conditions unattainable on Earth. 

Searches by LIGO scientific collaboration and the Virgo collaboration following GW170817 did not find any evidence for gravitational waves from a neutron star remnant \citep{AbEA2017h, Abbott:2018hgk, Abbott:2018wiz, LIGOScientific:2019eut}, although there has been a claim for evidence of short-duration post-merger signal~\citep{vanPutten:2018abw}. The nature of the remnant in a binary neutron star coalescence depends on the mass and spin of the remnant and the nuclear equation of state(for e.g. \citet{Baiotti:2016qnr, Piro:2017zec}). One possible outcome is the formation of a rapidly rotating, highly magnetized and long-lived ($t \geq 10$ s) massive neutron star (NS). Although no conclusive evidence for a long-lived remnant was found following GW170817, observations of X-ray afterglows of short gamma-ray bursts support this evolutionary pathway for a relatively large fraction of mergers \citep[e.g.,][]{RoOb2013}. Observations of GWs from  a long-lived post-merger remnant could help probe the complex physics governing the pre and post-merger phase, as well as help constrain the equation of state of massive remnants. Some predictions of GW signals from long-lived remnants suggest they may be observable with second-generation observatories out to $40$ Mpc (for e.g. ~\citet{DallOsso:2014hpa}), although more realistic analyses that account for the energy budget~\citep{Sarin:2018vsi} are more pessimistic and suggest that they might only be detectable with third-generation GW detectors~\cite{Punturo_2010, Hild_2011}.

There has been much work in exploring the GW emission from newly born magnetars. The nature of these GWs can depend sensitively on a number of aspects of neutron-star physics, including early cooling before transition to superfluidity, the effect of the magnetic field on the equilibrium shape, the internal dynamical state of a fully degenerate, oblique rotator, and the strength of the electromagnetic torque on the newly-born NS (e.g.~\citet{Cu2002,DallOsso:2014hpa,DoKo15,LaGl16}). The amplitudes and phases of the GWs depend on the complicated details of these physical mechanisms, which makes modeling and hence astrophysical inference from GW detections difficult. While there exists unmodeled Bayesian inference pipelines like Bayeswave~\citep{Cornish:2014kda, Chatziioannou:2017ixj} - which fits any signal using a wavelet expansion of variable dimensions - such analysis can be computationally expensive for long-transient signals considered in this paper. 

In this Letter, we develop methods for Bayesian inference of long-transient signals which are robust towards some modelling uncertainties. 
We focus on the phase of the signal in particular and derive phase-agnostic likelihoods which depend only on the spectral content of the signal. We use this likelihood in the context of Bayesian parameter estimation to constrain intrinsic properties of a (long-lived) remnant such as ellipticity and the braking index, using the millisecond magnetar model waveform~\citep{LaLe2017, Sarin:2018vsi} as an example waveform. We note that while we use this waveform model to study and develop parameter-estimation methods, we do not claim that this is a realistic model of long-lived post-merger emission. We show how this formalism performs both in the presence and the absence of a signal and how upper limits can be placed on gravitational-wave emission in the case of non-detection. 

\begin{figure}
 \includegraphics[width=3.5in]{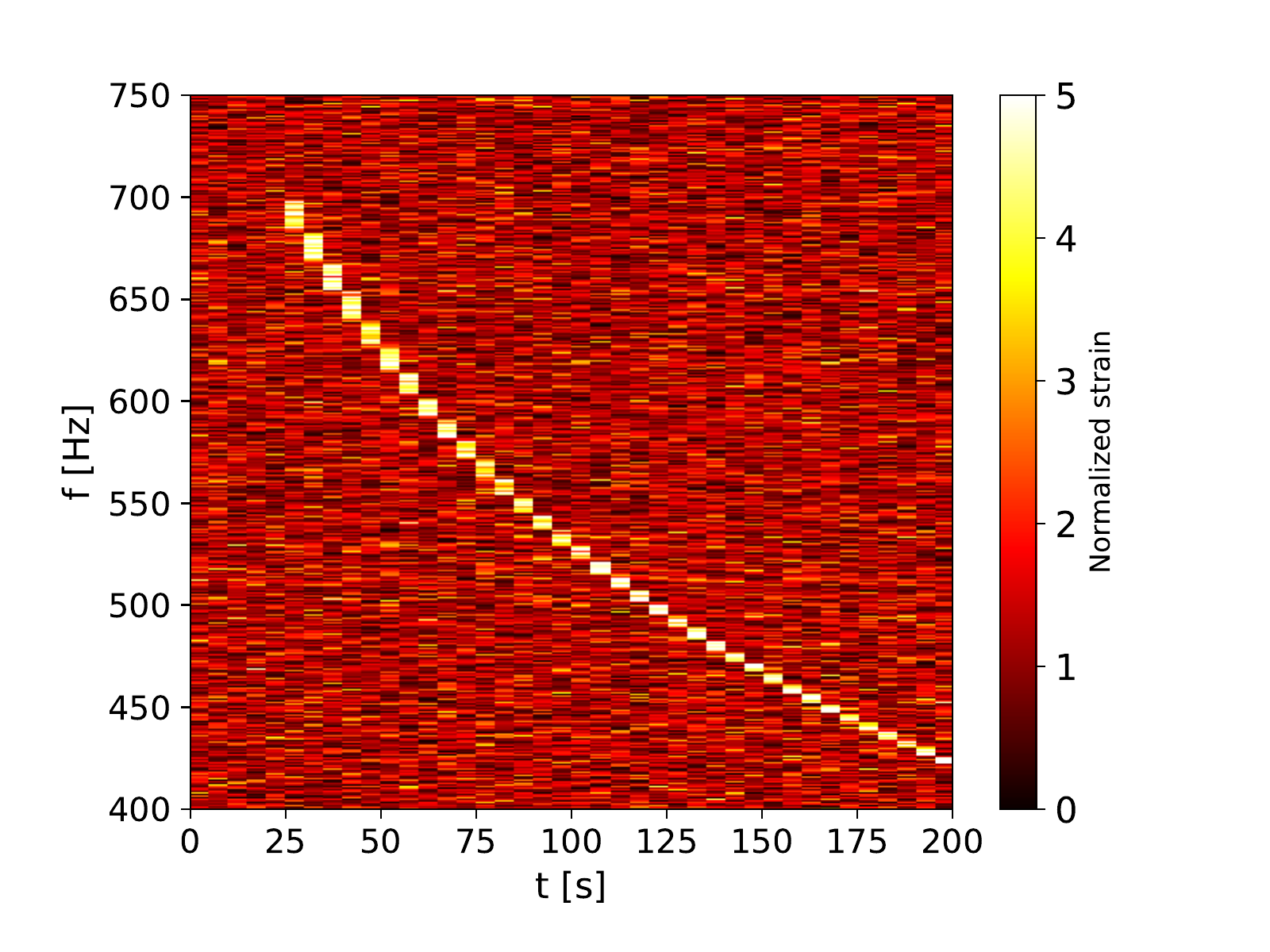}
 \caption{
   A normalized strain time-frequency map made with simulated Gaussian data recolored with O2 noise. A loud signal has been added for demonstration. The duration of each fast Fourier transform is 4 seconds and the entire map is 200 seconds long.}
 \label{fig:map}
\end{figure}

\section{Millisecond Magnetar Model}
The search for post-merger emission from GW170817 by LIGO and Virgo \citep{AbEA2017h, Abbott:2018hgk} considered a variety of possible signals, ranging from sub-second to hour-long timescales. In particular, the search for a signal from a long-lived remnant was based on a model derived from the dynamics of a spinning down neutron stars proposed by \citet{Sarin:2018vsi} and \citet{LaLe2017}. This model - hear-after referred to as the millisecond magnetar model -  derives the frequency evolution of the waveform from a spinning-down nascent neutron star with an arbitrary but fixed braking index $n$. We reproduce some of the details of the model below.

We assume that the rotational evolution of the star is described by the torque equation: $\dot{\Omega}\propto\Omega^n$, where $\Omega$ is the star's angular frequency. We also assume quadrupole GW emission caused by a non-zero ellipticity of the neutron star, so that $f(t) = \Omega (t) / \pi$. Integrating the torque equation yields the GW frequency:
\begin{equation}
	f(t)=f_0\left(1+\frac{t - t_0}{\tau}\right)^{1/(1-n)}, \;\;\; t \geq t_0.
\label{eq:lasky}
\end{equation}
Here $t_0$ is the start time of the emission (with some definition of $t=0$), $f_0$ is the initial GW frequency (at $t = t_0$) and $\tau$ is the spin-down timescale. Equation~\ref{eq:lasky} can describe emission from a variety of physical processes responsible for spin-down. For example, $n = 3$ describes magnetic dipole powered spin-down in vaccum, while $n = 5$ describes spin-down powered by emission of quadrupolar gravitational waves. The amplitude of the GW signal decreases with time as

\begin{equation}
h(t) = h_0 \left (1 + \frac{t - t_0}{\tau} \right)^{2/(1-n) } ,
\label{eq:amplEq}
\end{equation}
where we define an amplitude parameter $h_0$ as
\begin{equation}
h_0 = \frac{4\pi^2 G}{c^4} \frac{I_{zz}\epsilon}{d} f_0^2.
\label{eg:def_ampl}
\end{equation}
Here, $d$ is the distance of the source, $I_{zz}$ is the moment of inertia and $\epsilon$ is the eccentricity of the neutron star.

\begin{figure*}
\centering
\subfigure[]
{
    \scalebox{0.215}{\includegraphics{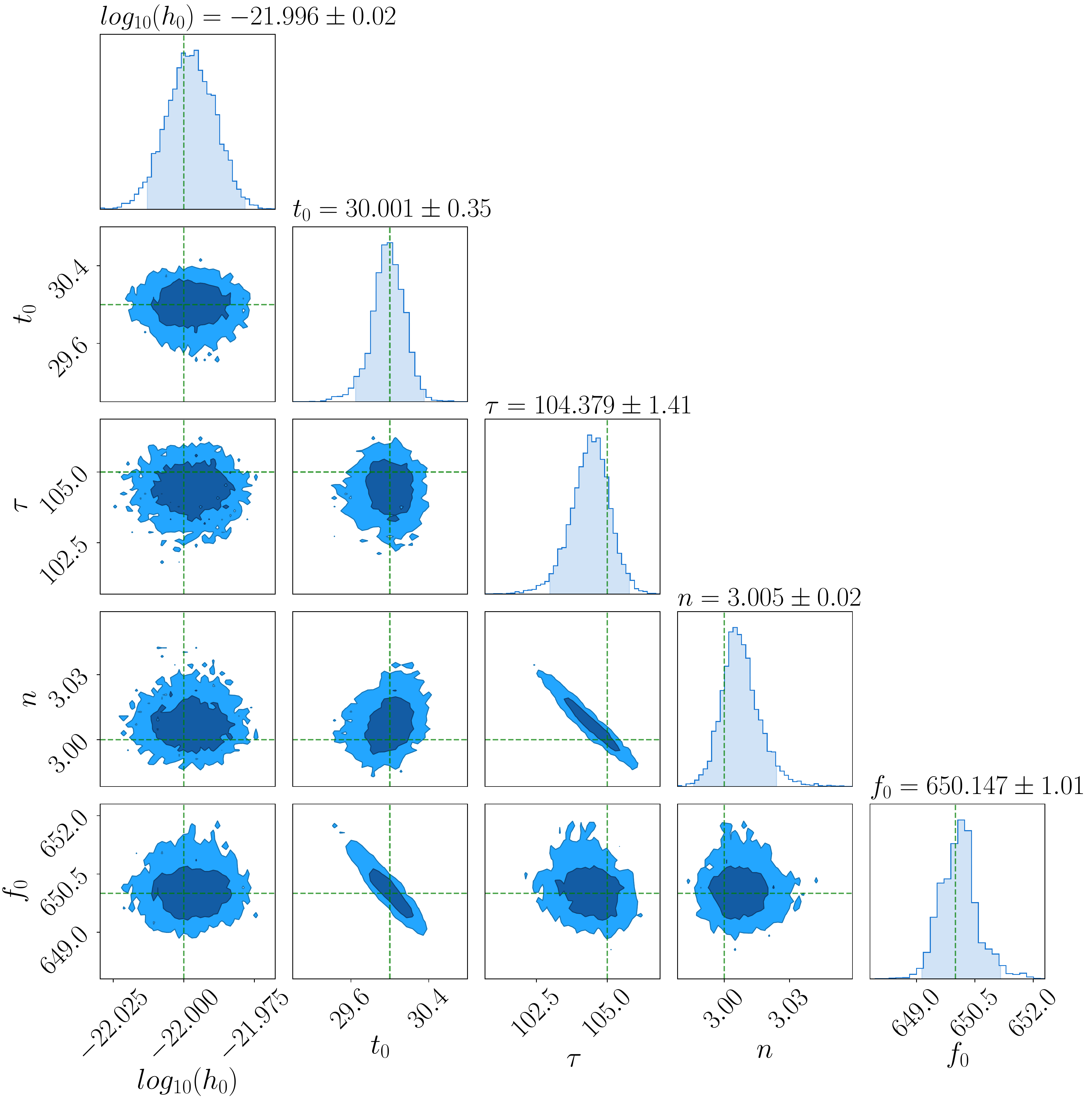}}
}
\subfigure[]
{
    \scalebox{0.215}{\includegraphics{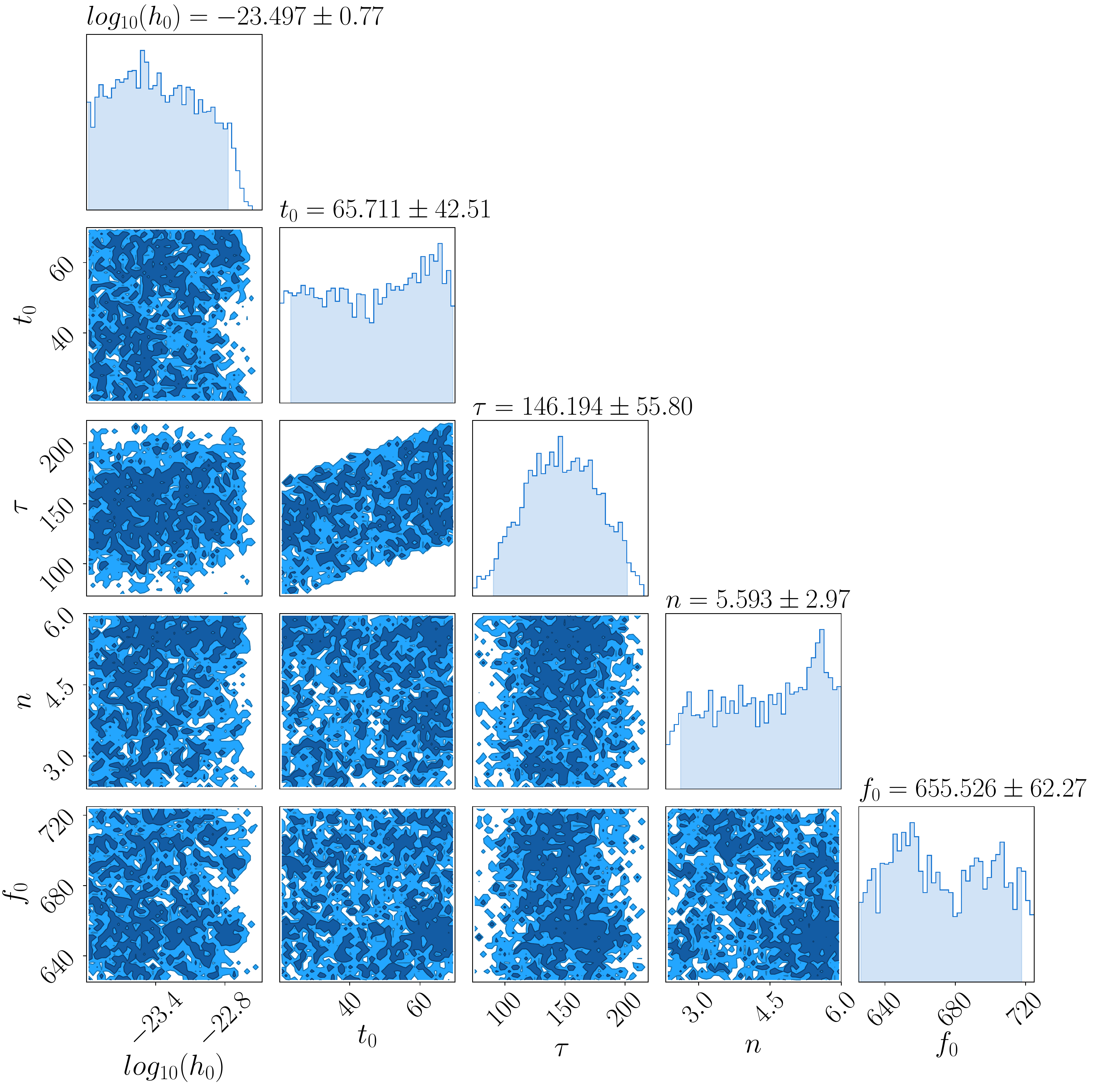}}
}
 \caption{Posteriors for a millisecond magnetar model simulation based on Eq~\ref{eq:lasky}. The colored regions in the 1-d posteriors show 95\% confidence intervals. The vertical green lines in the left panel are the true values corresponding to $log_{10}(h_0) = -22.0, t_0 = 30\,s, \tau =  105\,s, n = 3$ and $f_ 0 = 650$ Hz from left to right. The dark and the light regions in the 2-d posteriors are 68\% and 95\% confidence levels respectively. The posteriors on the right for Gaussian noise when no signal is present.  }
 \label{fig:posteriors}
\end{figure*}

\section{Likelihood Model}
A common way to search for GW sources that are difficult to accurately model is to look for excess power in time-frequency representations (tf-maps) of GW detector data \citep{Anderson:2000yy,SuEA2010,KlYa2008,ThKa2011}. To detect GWs, the tf-maps are parsed by pattern-recognition algorithms looking for statistically significant clusters of pixels --- for example, seeded~\citep{Burstegard, khan:2009} and seedless~\citep{ThCo2013,ThCo2014,CoTh2014} clustering algorithms using predefined templates have been widely used in the past. 

In this paper, we use tf-maps of discrete (complex) Fourier transforms of the data, normalized by the noise power spectral density (PSD).  An example map with a loud simulated signal is shown in Figure~\ref{fig:map}. \redtext{We assume that the noise is Gaussian and stationary over the period of analysis, and only consider correlations which are diagonal in both frequency and time. While this is a reasonable approximation for \greentext{simulated advanced LIGO data used in this paper} (see Appendix.~\ref{Ap_sec:noi_cov}), this would not be true for real interferometric data, in which case alternate basis like Discrete wavelet transforms (for e.g see~\cite{Littenberg:2010gf}) might be a much more suitable choice.}

The likelihood model we start with assumes that the residual noise when the signal is subtracted from the data is colored Gaussian noise. The Gaussian likelihood for a pixel in tf-map is given by~\citep{ VeRa2014} \footnote{We note here that the correct normalization of the Gaussian likelihood function in the frequency domain should be proportional to $ \sigma^{-2}$, and not to $\sigma^{-1}$ like in real time domain data. This is because frequency domain noise is generally complex in which both the real and imaginary parts of the noise are independently Gaussian. See Appendix D of \citet{Romano:2016dpx} for a careful examination of this.}  :
\begin{equation}
    \mathcal{L} (\tilde{d}_{ij}| \boldsymbol{\theta}) = \frac{2}{\pi T S^n_{j}}\exp\left ( -\frac{2}{T}\frac{|\tilde{d}_{ij} - \tilde{h}_{ij} (\boldsymbol{\theta})|^2}{ S^n_{j}} \right ),
    \label{eq:basic_likelihood}
\end{equation}
where $i, j$ are indices for the pixel at the $i$-th time-segment and $j$-th frequency bin of the tf-map. The terms $\tilde{d}_{ij}$, $\tilde{h}_{ij}$ and $S^n_{j}$ are the Fourier transform of the data, the signal model and the noise PSD in the pixel $i,j$ respectively. The term $T$ is the duration of the data used for the Fourier transform, and $\boldsymbol{\theta}$ is the vector of model parameters. We point to \citet{Thrane:2018qnx} for a review of methods of Bayesian inference used in gravitational-wave astrophysics. 

Given the uncertainty in the physics of the post-merger model describing the phase evolution of the remnants, we do not expect the signal to be accurate. Therefore we need to incorporate our ignorance of the true phase of the signal when analyzing the data. One way to do this is by marginalizing the phase of each pixel independently of other pixels\footnote{ The phase marginalization being done here is different from the one used in parameter estimation analysis of compact binary coalescence , e.g \citet{VeRa2014}. The phase evolution of compact binary waveforms is well understood, and it is only the initial phase which is marginalized over.}. The resultant phase marginalized likelihood depends only on the spectral content of the signal, and can be written as
\begin{equation}
\begin{split}
        \mathcal{L}_{\phi} (\tilde{d}_{ij}| \bar{\theta}) = \frac{2}{\pi T S^n_{j}}\exp \left( -\frac{2}{T}\frac{|\tilde{d}_{ij}|^2 + |\tilde{h}_{ij}|^2}{ S^n_{j}}\right)  \times \\ I_0 \left [\frac{4}{T} \frac{|\tilde{d}_{ij}||\tilde{h}_{ij}|}{S^n_{j}}  \right ], 
\end{split}
\label{eq:Bessel_likelihood}
\end{equation}

\noindent where $I_0(x)$ is the zeroth-order modified Bessel function of the first kind. We refer the reader to \redtext{Appendix.~\ref{Ap_sec:Bessel_like}} for the derivation of Eq.~\ref{eq:Bessel_likelihood}. This likelihood is for a single pixel of one interferometer. We take the product of likelihoods over all pixels to extend it over the entire tf-map. The simplest way to incorporate multiple detectors is to take the product of likelihoods for each detector:

\begin{equation}
    \mathcal{L} ( \{ d_k\} | \boldsymbol{\theta}) = \prod_{i,j, k} \mathcal{L}_{\phi} (  d_{i j k} | \boldsymbol{\theta}),
\end{equation}

\noindent where $k$ is an index over interferometers.

Tests of the likelihood in Eq.~\ref{eq:Bessel_likelihood} show that the recovered parameters suffer from biases unless the exact spectrum of the noise is known. A common way to estimate the noise PSD is by calculating the mean of the PSDs of neighbouring or off source data segments. This estimate has a variance about the true PSD of the noise, which would need to be accounted for when large amounts of data are analyzed. One way to do this is to marginalize over the true PSD in a pixel given our measurement of $S^n_{j}$. Starting with the Gaussian likelihood in Eq.~\ref{eq:basic_likelihood} and using a $\chi^2$ prior for the true PSD gives a likelihood based on a Student-t distribution for each pixel:
\begin{equation}
    \mathcal{L}_S (\tilde{d}_{ij}| \bar{\theta}) = \frac{4 \,\Gamma\left (1 + \frac{\nu}{2}\right )}{\pi T \nu S^n_{j} \,\Gamma(\nu/2)  } \left [ 1 + \frac{4}{T} \frac{|\tilde{d}_{ij} - \tilde{h}_{ij}|^2}{\nu \, S^n_{j}}  \right]^{-\left(1 + \frac{\nu}{2}\right )}.
    \label{eq:student_t}
\end{equation}
Here $\nu$ is the number of degrees of freedom of the $\chi^2$ prior. A natural value for $\nu$ is $\nu = 2N$, where $N$ is the number of data segments used to calculate $S^n_{j}$. As pointed out in \citet{Rover:2008yp}, \greentext{Student-t distributions with fewer degrees of freedom have larger tails, implying that they better account for uncertainties in the noise PSD and yield more robust inferences. This is however limited by the assumption that the noise is stationary. Methods which simultaneously model the noise along with the signal such as in ~\cite{ Littenberg:2013gja} and ~\cite{ Cornish:2014kda} would work better for non-stationary real data.}   We find that using fewer degrees of freedom gives better inferences; we estimate the PSD using $N = 40$ segments, and use $\nu = N$ in all the examples shown in this paper.

\begin{figure*}
    \centering
    \includegraphics[width=6.5in]{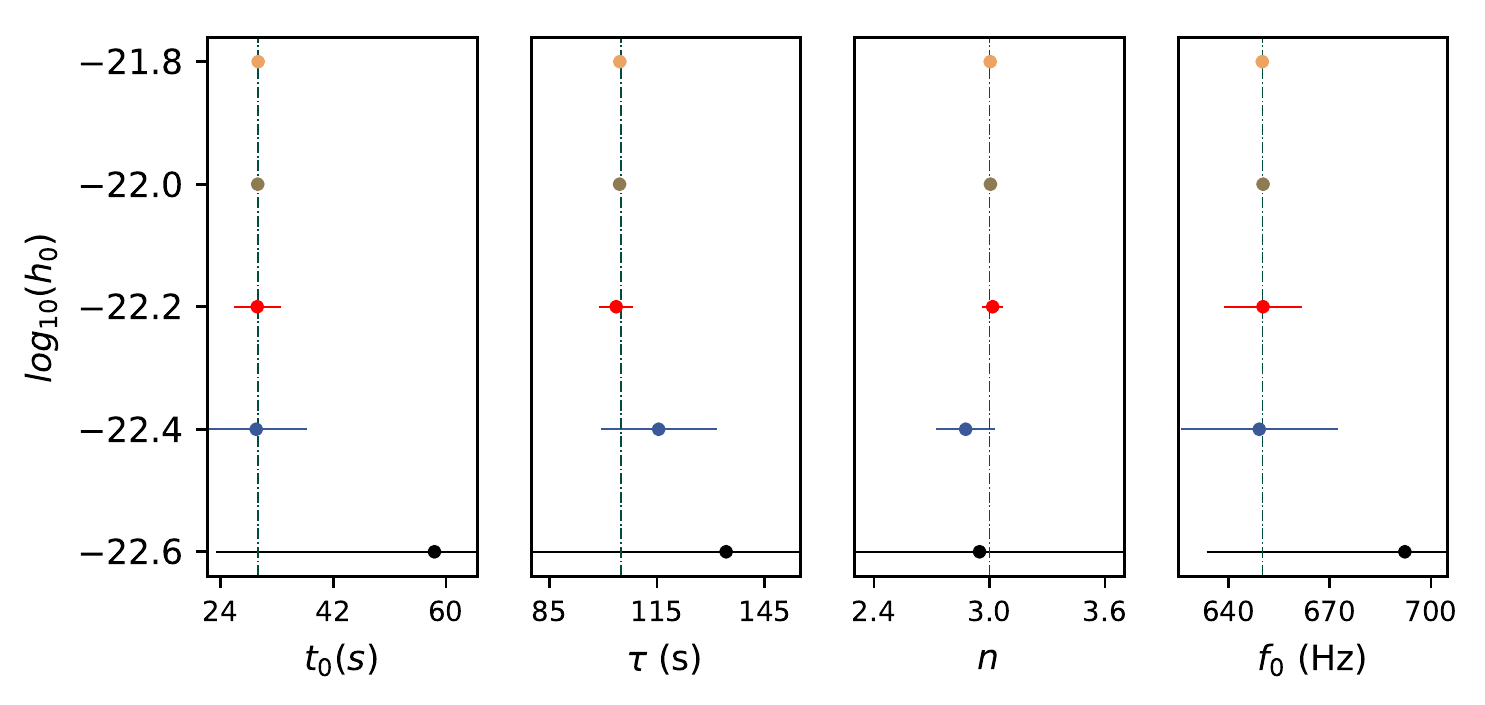}
    \caption{Posterior recoveries for the millisecond magnetar model showing the levels at which spectral parameters are constrained at different amplitude values. The vertical axis on the left shows the amplitude values used for the simulations. The solid dots are the maximum a posteriori values and the error bars correspond to 95\% confidence levels.  The vertical dashed-dotted line are the true values. Note that the x axis for these plots do not show to the full prior range and we have zoomed in to see the error bars better. }
    \label{fig:vary_ampls_test}
\end{figure*}

Having corrected for the PSD variance, we marginalize over the phase of the signal tf-map again to obtain a likelihood functions based on hyper-geometric functions for each pixel:
\begin{multline}
     \mathcal{L}_{S, \phi} (\tilde{d}_{ij}| \bar{\theta}) = \frac{\alpha_{ij}}{2} (1 - \beta_{ij})^\gamma \prescript{}{2}{F}_1 \left(0.5, -\gamma, 1, \frac{2 \beta_{ij}}{\beta_{ij}-1} \right) + \\ \frac{\alpha_{ij}}{2} (1 + \beta_{ij})^\gamma \prescript{}{2}{F}_1 \left(0.5, -\gamma, 1, \frac{2 \beta_{ij}}{\beta_{ij}+1} \right)  ,
\label{eq:hypergeo_likelihood}
\end{multline}
where, 
\begin{equation}
\begin{split}
    \alpha_{ij} = \frac{2 \,\Gamma\left (1 + \frac{\nu}{2}\right )}{\pi \nu S^n_{j} \,\Gamma(\nu/2)  } \left [  \frac{\nu \, S^n_{j} + | \tilde{d}_{ij} |^2 + |\tilde{h}_{ij}|^2}{\nu \, S^n_{j}}  \right]^{-\left(1 + \frac{\nu}{2}\right )},\\ \\
        \beta_{ij} = \frac{2\, |d_{ij}| |h_{ij}|}{|d_{ij}|^2 + |h_{ij}|^2 + \nu S^n_{j}} \, , \, \gamma = -\left(1 + \frac{\nu}{2}\right ) .
\end{split}
\label{eq:def_alpha_beta_gamma}
\end{equation}

We point the reader to \redtext{ Appendix.~\ref{Ap_sec:hypergeo_like}} for the derivation and more details about both Eq.~\ref{eq:student_t} and Eq.~\ref{eq:hypergeo_likelihood}.

\section{Analysis}
\label{Sec:Analysis}
We now use the likelihood in Eq.~\ref{eq:hypergeo_likelihood} to recover a simulated signal from the millisecond magnetar model added to Gaussian noise colored with the O2 PSD of Hanford and Livingston Advanced LIGO detectors. We make tf-maps which are 200 seconds long, divided into 4\,s Tukey-windowed FFT pixels. In this analysis, we assume that we know the distance $d$ and the sky-location of the remnant, which were simulated to be the same as GW170817 and its electromagnetic counterpart \citep{AbEA2017b, GBM:2017lvd}, i.e (ra, decl)  =  ($13.1634$  Hrs, $-23.3185^\circ$) and $d = 40$ Mpc. \redtext{We also assume a polarization angle of $\psi$ = 0 and an optimal orientation of the remnant, i.e $\cos \iota = 1$}

We sample over the five-dimensional parameter space $\boldsymbol{\theta} = \{ h_0, t_0, \tau, n,  f_0 \}$ using {\sc PyMultiNest} \citep{BuGe2014}, a python wrapper for the Nested Sampling implementation of {\sc MultiNest} \citep{FeHo2009}. We use flat priors for all parameters \footnote{The priors extend from $20s$ to $70s$ for $t_0$,from $2.3$ to $5$ on $n$, and from $625$ Hz to $725$ Hz on $f_0$. The parameter $\tau$ is degenerate with $t_0$,  so in place of $\tau$ we actually sample over $T = \tau + t_0$ with a flat prior between $50$s to $150$s.} except $h_0$, for which we use a uniform in log prior from $10^{-24}$ to $10^{-21}$. The left panel of Figure~\ref{fig:posteriors} shows an example of the parameter estimation of a simulated signal. In this case, the parameters are constrained roughly to a percent level.

The right panel of Figure~\ref{fig:posteriors} shows results from an analysis with Gaussian noise. In the absence of a signal, the posterior of the signal amplitude $h_0$ can be used to place upper limits on some properties of the remnant. Here, we get a 95\% upper limit on $h_0$ of $2.1 \times 10^{-23}$ with a uniform in log prior. Using the posterior samples and with Eq.~\ref{eg:def_ampl}, we can constrain the physical parameters of the remnants. In this case for example assuming a distance of 40\,Mpc and the same fiducial moment of inertia as in Ref.~\citep{Abbott:2018hgk} of $I_{zz} = 4.34 \times 10^{38} \, \si{kg.m^2}$,  we get a 95\% limit on ellipticity of 0.499. In reality we might not know the distance and the sky-position in the absence of an electromagnetic counterpart, and the moment of inertia, \redtext{polarization and inclination angle} of the remnant would also not be known precisely. These extra sources of uncertainty would need to be folded into both the analysis and the upper-limit calculation, either as extra parameters or using constraints from other measurements (for e.g. the distance measurement from the inspiral signal). 

The upper limit on $h_0$ is also consistent with Figure~\ref{fig:vary_ampls_test}, where we attempt to recover the simulated signals different amplitude levels while keeping constant the spectral parameters. The figure shows $95\%$ confidence intervals with which the spectral parameters are recovered at different amplitudes. While the posteriors are well constrained for $h_0 \geq 4.0 \times 10^{-23}$, for a signal with amplitude  $h_0 \leq 2.5 \times 10^{-23}$, the posteriors span almost the entire prior range. 

\begin{figure}
\includegraphics[width=3.5in]{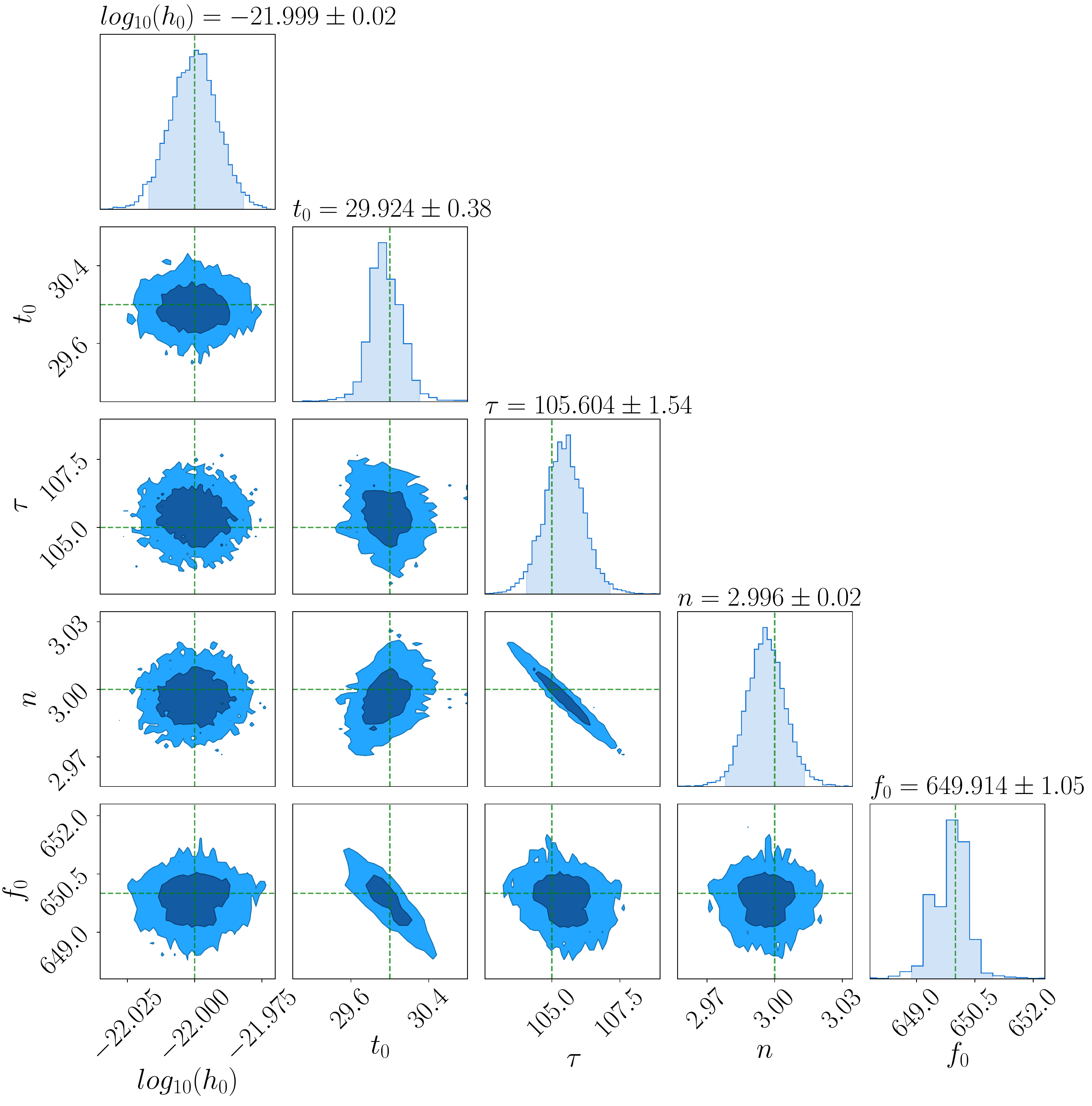}
\caption{ Posteriors recoveries for a simulation with small fluctuations added to the phase evolution. The colored regions in the 1-d posteriors show 95\% confidence intervals. The vertical green lines in the left panel are the true values corresponding to $log_{10}(h_0) = -22.0, t_0 = 30\,s, \tau =  105\,s, n = 3$ and $f_ 0 = 650$ Hz from left to right. The dark and the light regions in the 2-d posteriors are 68\% and 95\% confidence levels respectively.}
\label{fig:phase-independence}
\end{figure}

We finally use the marginalized likelihood model on simulated signals with incorrect phase evolution models. We first test this in the frequency-domain with phase-scrambled maps --- which are tf-maps with random fluctuations added to the phase of each pixel. We find that the recovered posteriors are consistent with the true values, as expected for this statistic. We also perform a similar test by adding fluctuations to the time-domain phase evolution of the signal. We note that care is needed in the time-domain that the fluctuations not be large enough to affect the frequency evolution in the signal. The Figure~\ref{fig:phase-independence} show the results for a signal with small fluctuations added in the time-domain, which demonstrate posterior recoveries consistent with the true parameters of the model.  

\section{Conclusion}
Post-merger signals from neutron stars are a promising source of GWs for second and third generation gravitational-wave detectors. In this paper, we have described the application of a Bayesian likelihood formalism to the characterization of long-duration post-merger signals from binary neutron star mergers. We showed that this formalism is robust against fluctuations in phase evolution and is capable of constraining and measuring important astrophysical parameters like the spin, the braking index, moment of inertia and the eccentricity of magnetars. 
We note in particular the possibility to estimate the braking index of the remnant NS with gravitational-wave data. There have been only two measurements of braking indices of millisecond magnetars to date using, x-ray observations following short gamma-ray bursts \citep{LaLe2017}. Braking index measurements would be of particular interest since that gives direct information of the underlying mechanics of the spin-down. In conjugation with on-going developments~\citep{Takami:2014zpa, Bernuzzi:2015rla, Tsang:2019esi} in modeling of post-merger GW emission, parameter estimation methods can also help constrain the nuclear equation of state at very high densities. 

As the second-generation gravitational-wave detectors progress towards their design sensitivity, it is plausible that there will be a detection of a long-transient gravitational-wave signal in the coming observing runs. In addition to post-merger searches, analyses of these signals also benefit from development of parameter estimations methods which make minimal model assumptions. We are planning further developments of robust methods of sky localization for transients which is especially important. One assumption we have made throughout this work is that the spectral model is well known. We are also developing parameter estimation methods which can handle a wider range of model uncertainties, and help in extracting astrophysical information from future detections. 

\section*{Acknowledgements}
We are grateful to Ling Sun, Joe Romano, David Keitel and Paul Schale and for useful discussion and comments. We also thank Tyson Littenberg for insightful comments and suggestions that improved the paperSB acknowledges support by the Hoff Lu Fellowship at the university of Minnesota,  and by NSF grant PHY-1806630. MC is supported by the David and Ellen Lee Postdoctoral Fellowship at the California Institute of Technology. PDL is supported by ARC Future Fellowship FT160100112 and ARC Discovery Project DP180103155. ET and CT are supported by CE170100004. ET is supported by FT150100281. The authors are thankful for the computing resources provided by LIGO Laboratory. LIGO was constructed by the California Institute of Technology and Massachusetts Institute of Technology with funding from the National Science Foundation and operates under cooperative agreement PHY-0757058. All posterior corner plots were made with ChainConsumer~\citep{ChainConsumer}. The code for the analysis in this paper is available upon request. A public release is being planned for the near future. This paper has been assigned document number LIGO-P1900107.

\appendix
\section{Phase Marginalized likelihood}
\label{Ap_sec:Bessel_like}

We derive Eq.~\ref{eq:Bessel_likelihood}
 the phase marginalized Bessel function likelihood. We begin with Eq.~\ref{eq:basic_likelihood} and write it by explicitly separating the phase term as,

\begin{multline}
    \mathcal{L} (\tilde{d}_{ij}| \bar{\theta}, \phi^h_{ij}) = \frac{2}{\pi T S^n_{j}}\exp\left ( -\frac{2}{T}\frac{|\tilde{d}_{ij}|^2 + |\tilde{h}_{ij}|^2}{ S^n_{j}} \right ) \times \\ \exp\left ( \frac{4}{T} \frac{|\tilde{d}_{ij}||\tilde{h}_{ij}| \cos({\phi^d_{ij} - \phi^h_{ij} }) }{S^n_{j}} \right ),
    \label{eq:unmarg}
\end{multline}

\noindent where $\phi^h_{ij}$ and $\phi^d_{ij}$ are the model and data phase in the pixel $i, j$. We marginalize over $\phi^h_{ij}$ (or over $\phi_{ij} = \phi^d_{ij} - \phi^h_{ij}$) with a uniform prior as a natural choice. The marginalization integral is,

\begin{equation}
    I^{\phi}_{ij} = \frac{1}{2 \pi} \int_0^{2 \pi} d \phi \exp\left (  \frac{4}{T} \frac{|\tilde{d}_{ij}||\tilde{h}_{ij}|}{S^n_{j}} \cos{\phi } \right ).
\end{equation}

The integral can be described in terms of a zeroth-order modified Bessel function of the first kind \citep{ArfkenGeorge1972MMfP}, such the marginalized likelihood is, 

\begin{equation}
\begin{split}
        \mathcal{L}_{\phi} (\tilde{d}_{ij}| \bar{\theta}) = \frac{2}{\pi T S^n_{j}}\exp \left( -\frac{2}{T}\frac{|\tilde{d}_{ij}|^2 + |\tilde{h}_{ij}|^2}{ S^n_{j}}\right)  \times \\ I_0 \left [\frac{4}{T} \frac{|\tilde{d}_{ij}||\tilde{h}_{ij}|}{S^n_{j}}  \right ].
\end{split}
\end{equation}

\section{Phase and PSD Marginalized likelihood}
\label{Ap_sec:hypergeo_like}
A common way to estimate the noise PSD at some frequency is by averaging over the PSDs estimate from $N$ neighbouring time segments:

\begin{equation}
    S_{avg} = \frac{1}{N} \sum_i^N S_i.
    \label{eq:PSD_estimator}
\end{equation}

In the frequency domain, both the real and imaginary parts of the noise are assumed to be drawn from a colored Gaussian noise. If the true PSD is $S$, then the expectation value of both the real and imaginary part of the noise is $S/2$. Then the following sum follows a $\chi^2$ distribution with $2N$ degrees of freedom:
\begin{equation}
 \frac{2}{S} \sum_i^N S_i = \frac{2N}{S} S_{avg}. 
\end{equation}

In general if some data $X \sim N(0, \sigma^2)$ and $ \nu Y^2/\sigma^2 \sim \chi^2_{\nu}$, where $Y$ is an estimator for $\sigma$, then the random variable $t = X/Y$ will form a Student-t distribution with $\nu$ degrees of freedom \citep{Stat_methods}. Using that here, we get the PSD marginalized likelihood for a pixel: \\

\begin{equation}
    \mathcal{L}_S (\tilde{d}_{ij}| \bar{\theta}) = \frac{4 \,\Gamma\left (1 + \frac{\nu}{2}\right )}{\pi T \nu S^n_{j} \,\Gamma(\nu/2)  } \left [ 1 + \frac{4}{T} \frac{|\tilde{d}_{ij} - \tilde{h}_{ij}|^2}{\nu \, S^n_{j}}  \right]^{-\left(1 + \frac{\nu}{2}\right )},
\end{equation}

\noindent with the natural choice of $\nu = 2N$. Note that since we start with the complex Gaussian distribution  Eq.~\ref{eq:basic_likelihood}, the exponent is not $-(\nu +1)/2$
and so this is not an exact Student-t distribution in $|\tilde{d}_{ij} - \tilde{h}_{ij}|$. Now we marginalize over the phase. We define the $\alpha_{ij}$,$\beta_{ij}$ and $\gamma$ variables as in Eq.~\ref{eq:def_alpha_beta_gamma} which allows us to write the likelihood as,

\begin{equation}
    \mathcal{L}_S^{ij} (\tilde{d}_{ij}| \bar{\theta}) = \alpha_{ij} \left [1 - \beta_{ij} \cos ({\phi^s_{ij} - \phi^h_{ij} }) \right]^{\gamma}.
\end{equation}

We now marginalize over the phase term $\phi_{ij} =  {\phi^s_{ij} - \phi^h_{ij} }$;

\begin{equation}
      \mathcal{L}_S^{ij} (\tilde{d}_{ij}| \bar{\theta}) = \frac{\alpha_{ij}}{2 \pi} \int_0^{2 \pi} d \phi_{ij} \left [1 - \beta_{ij} \cos \phi_{ij} \right]^{\gamma}
\end{equation}

\noindent The integral can be written in terms of Gauss hypergeometric functions as, 

\begin{multline}
         \mathcal{L}_S^{ij} (\tilde{d}_{ij}| \bar{\theta}) = \frac{\alpha_{ij}}{2} (1 - \beta_{ij})^\gamma \prescript{}{2}{F}_1 \left(0.5, -\gamma, 1, \frac{2 \beta_{ij}}{\beta_{ij}-1} \right) + \\ \frac{\alpha_{ij}}{2} (1 + \beta_{ij})^\gamma \prescript{}{2}{F}_1 \left(0.5, -\gamma, 1, \frac{2 \beta_{ij}}{\beta_{ij}+1} \right),   
\end{multline}

\noindent which gives the phase and PSD marginalized likelihood for each pixel.

\section{The noise covariance of the tf-map }
\label{Ap_sec:noi_cov}

\begin{figure*}
\includegraphics[scale=0.4]{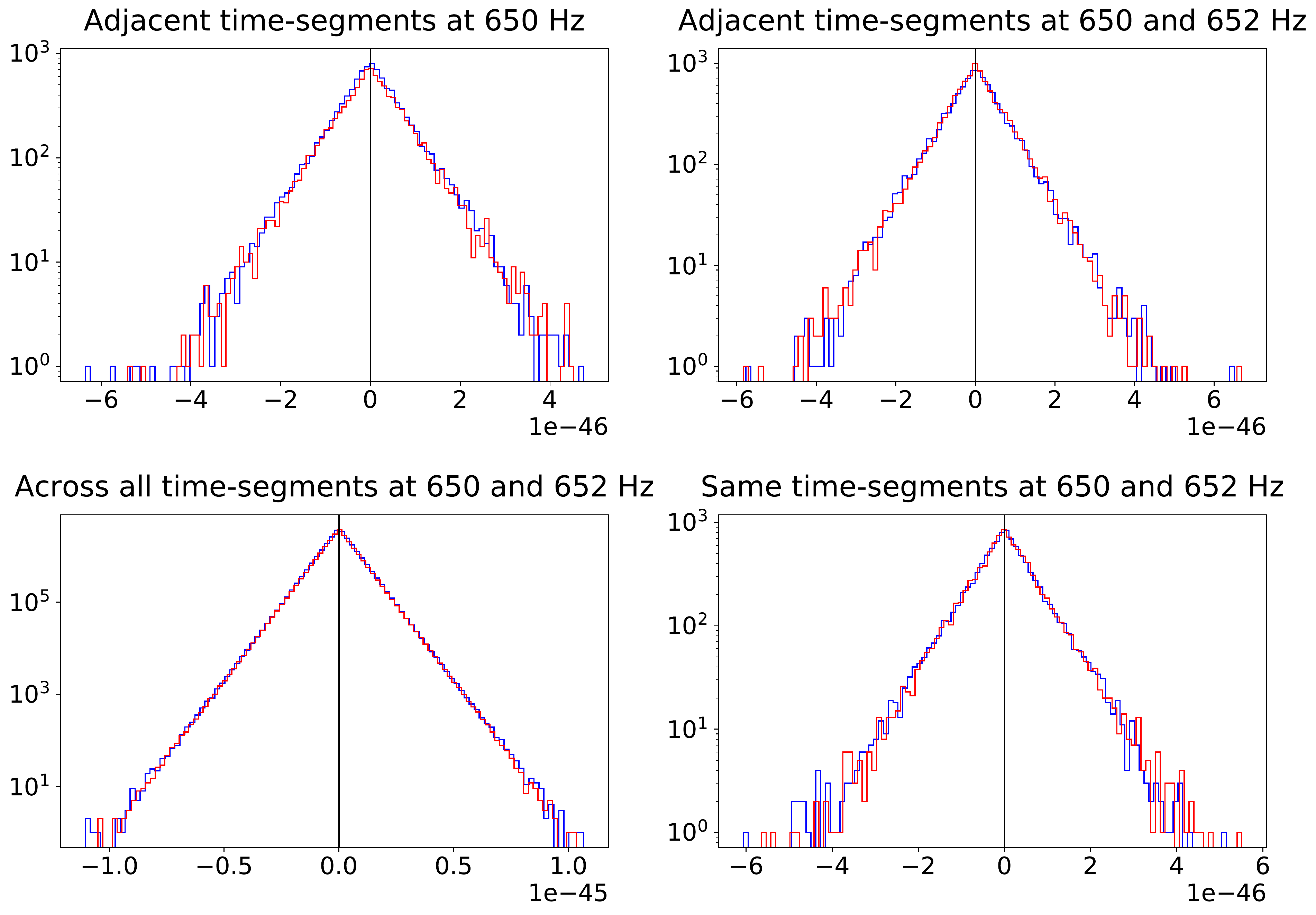}
\caption{Starting from the top left in clockwise direction, the plots show histograms of (i) Correlations between two frequency bins at 650 Hz at adjacent time-segments, (ii) Correlations between bins at 650 and 652 Hz at adjacent time-segments, (iii) Correlations between bins at 650 and 652 Hz across all time-segments and  (iv) Correlations between bins at 650 and 652 Hz at the same time-segments. The blue and the red traces correspond to the real and imaginary parts of the correlations. The absolute values of the means of the correlations are (clockwise from top-left) $3.3 \times 10^{-49}, 8.8 \times 10^{-49}, 6.5 \times 10^{-52}, 1.7 \times 10^{-48}$. }
\label{fig:Noi_corr}
\end{figure*}

\redtext{
In this section we study the covariance of the simulated noise used in this paper, and test the assumption of only using correlations diagonal in frequency and time in the likelihood. 

For the first test we simulated $\mathcal{O}(10,000)$ seconds of stationary advanced LIGO colored Gaussian noise and created tf-maps with 4 second tukey-windowed FFTs as in Sec.~\ref{Sec:Analysis}. We then made histograms measuring different types of correlations using these FFTs which we list below:

\begin{itemize}
    \item Correlations between bins of the same frequency but at neighbouring time-segments - i.e time-segment adjacent to each other in the tf-map
    \item Correlations between distinct frequency bins at neighbouring time-segments 
    \item Correlations between frequency bins both distinct and at same frequencies, across all time-segments (i.e not just neighbouring ones)
    \item Finally for completeness, distinct frequency bins at the same time-segments 
\end{itemize}

 \greentext{Example histograms are plotted in Fig.~\ref{fig:Noi_corr} for these different types of correlations.} We see that the mean values of the noise-correlation histograms is consistent with zero. This is expected when any correlations between these frequency bins across time-segments is negligible. This was true across different frequency choices in the \greentext{advanced LIGO sensitive frequency band from $\sim 20 - 1000$ Hz.}

As another test we studied correlations across time using the time-domain auto-correlation function. We use noise correlation duration as a metric to measure the duration of correlations of a stationary (ergodic) random data. It is defined as~\citep{Bendat&Piersol:20003ed}:

\begin{equation}
    T_n = \frac{2}{R_{xx}(0)} \int_0^{\infty} |R_{xx}(\tau)| \, d \tau,
\end{equation}

where $R_{xx}(\tau)$ is the time-domain auto-correlation function calculated as the inverse Fourier transform of a two-sided PSD. When using the O2 PSD of advanced LIGO without narrow-band features, and with a sampling frequency of 2048 Hz ( $f_{max} = 1024$ Hz) we get $T_n \approx 0.34$ seconds. This is already more than an order of magnitude smaller than the segment duration being used in this paper, but most of the correlations comes from the lowest frequencies which are not typically used in gravitational-wave analysis. Indeed if we only consider frequencies above 30 Hz the correlation duration drops to around 0.02 seconds.  

\begin{figure}
    \includegraphics[scale=0.28]{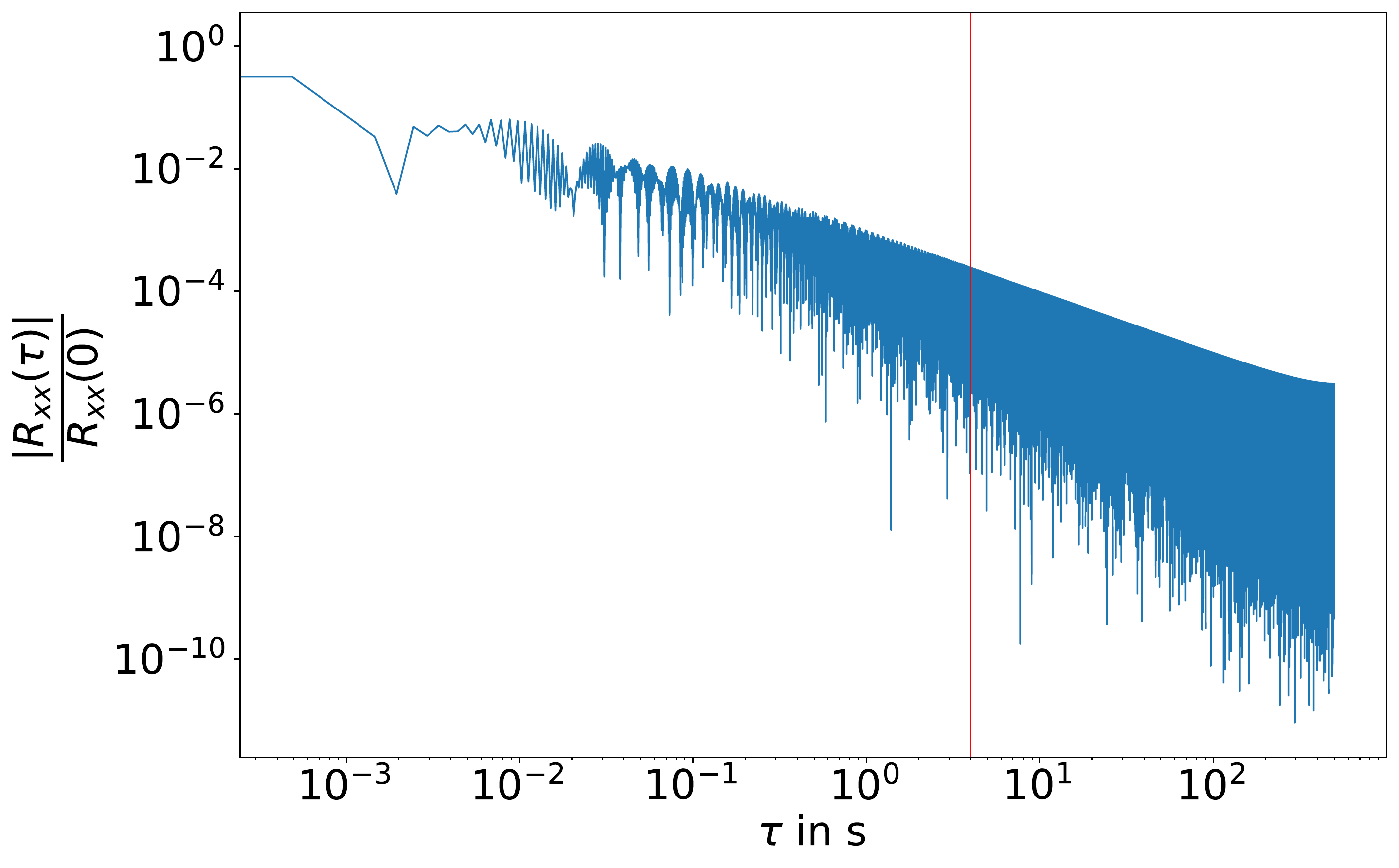}
    \caption{The absolute value of the auto-correlation function is plotted as a function of the time-difference. The correlation value at four seconds - indicated by the red line - is atleast two orders of magnitude smaller than 1. }
    \label{fig:auto_corr}
\end{figure}

Finally we plot the absolute value of the auto-correlation function which was computed using an IFFT of the PSD in Fig.~\ref{fig:auto_corr} with a low frequency cut-off of 30 Hz. This shows that the correlations at $\tau = 4$ seconds are several orders of magnitude smaller in magnitude than $R_{xx}(\tau = 0)$. We should stress again that these tests are valid only for noise which is stationary and Gaussian for the duration of the entire spectrogram without strong narrow-band features; assumptions which would not generally be true with real data.}

\bibliography{references}
\bibliographystyle{mnras}

\bsp
\label{lastpage}
\end{document}